\documentclass[aps,prb,reprint,superscriptaddress]{revtex4-2}
\usepackage{graphicx}
\usepackage{amsmath}
\usepackage{amssymb}
\usepackage{color}
\usepackage{hyperref}
\usepackage{siunitx}
\usepackage{easyReview}

\begin{document}
\title{Unusual magnetic and transport properties in the Zintl phase Eu$_{11}$Zn$_6$As$_{12}$}
\author{Zhiyu Zhou}
\affiliation{Key Laboratory of Quantum Materials and Devices of Ministry of Education, School of Physics, Southeast University, Nanjing 211189, China}
\author{Ziwen Wang}
\affiliation{Key Laboratory of Quantum Materials and Devices of Ministry of Education, School of Physics, Southeast University, Nanjing 211189, China}
\author{Xiyu Chen}
\affiliation{Key Laboratory of Quantum Materials and Devices of Ministry of Education, School of Physics, Southeast University, Nanjing 211189, China}
\author{Jia-Yi Lu}
\affiliation{School of Physics, Interdisciplinary Center for Quantum Information and State Key Laboratory of Silicon and Advanced Semiconductor Materials, Zhejiang University, Hangzhou 310058, China}
\author{Junchao Zhang}
\affiliation{Key Laboratory of Quantum Materials and Devices of Ministry of Education, School of Physics, Southeast University, Nanjing 211189, China}
\author{Xiong Luo}
\affiliation{Key Laboratory of Quantum Materials and Devices of Ministry of Education, School of Physics, Southeast University, Nanjing 211189, China}
\author{Guang-Han Cao}
\affiliation{School of Physics, Interdisciplinary Center for Quantum Information and State Key Laboratory of Silicon and Advanced Semiconductor Materials, Zhejiang University, Hangzhou 310058, China}
\affiliation{Collaborative Innovation Centre of Advanced Microstructures, Nanjing University, Nanjing 210093, China}
\author{Shuai Dong}
\affiliation{Key Laboratory of Quantum Materials and Devices of Ministry of Education, School of Physics, Southeast University, Nanjing 211189, China}
\author{Zhi-Cheng Wang}
\email{wzc@seu.edu.cn}
\affiliation{Key Laboratory of Quantum Materials and Devices of Ministry of Education, School of Physics, Southeast University, Nanjing 211189, China}
\date{\today}

\begin{abstract}

Narrow-gap rare-earth Zintl phases frequently exhibit fascinating physical phenomena due to their various crystal structures, complex magnetic properties, and tunable transport behaviors. Here we report the synthesis, magnetic, thermodynamic, and transport properties of a Eu-containing Zintl arsenide, Eu$_{11}$Zn$_6$As$_{12}$, which consists of infinite chains of Eu cations and anionic frameworks constructed from corner-sharing ZnAs$_4$ tetrahedra. Eu$_{11}$Zn$_6$As$_{12}$ exhibits complicated magnetic behavior owing to intricate exchange interactions mediated by the discrete anionic fragments. Two long-range magnetic transitions at 22 K ($T_\mathrm{N}$) and 9 K ($T^*$), as well as exceptionally strong ferromagnetic fluctuations around 29 K ($T_\mathrm{F}$), are indicated by the susceptibility, heat capacity and resistivity measurements. Besides, Eu$_{11}$Zn$_6$As$_{12}$ displays metallic behavior, attributable to the hole carriers doped by slight Eu vacancies or the mixed valence of Eu$^{2+}$ and Eu$^{3+}$. A prominent resistivity peak occurs around $T_\mathrm{N}$, which is rapidly suppressed by the applied field, leading to a prominent negative magnetoresistance effect. A resistivity hysteresis is observed below 5 K, caused by a small net ferromagnetic component. Our study presents the distinct magnetic and transport properties of Eu$_{11}$Zn$_6$As$_{12}$, and further experiments are required to elucidate the origin of these novel behaviors. Moreover, our findings demonstrate that Eu-based Zintl phases are a fertile ground to study the interplay between magnetism and charge transport.

\end{abstract}
\maketitle

\section{Introduction}
Zintl phases, commonly known as valence-precise compounds consisting of electropositive cations and complex anionic frameworks, continue to attract significant research interest due to their diverse structures and physical properties, such as colossal magnetoresistance (CMR), thermoelectricity, anomalous Hall effect (AHE), and superconductivity~\cite{kauzlarich1996ChemistryStructureBonding,Fang2022a,devlinEu11Zn2018,Payne2001a,Shuai2017,chenZintlphaseEu2ZnSb2Promising2019,Cao2023}. Recently, multiple Eu-containing Zintl phases have been proposed as magnetic topological materials, including EuSn$_2$As$_2$, EuMg$_2$Bi$_2$, EuIn$_2$As$_2$, EuCd$_2$As$_2$, and Eu$_5$In$_2$Sb$_6$~\cite{Li2019a,Li2021a,Marshall2021a,Xu2019,Zhang2020a,huaDiracSemimetalTypeIV2018,maSpinFluctuationInduced2019,xuUnconventionalTransverseTransport2021,Rosa2020}. Among these, the CaAl$_2$Si$_2$-type EuCd$_2$As$_2$ and its sibling compounds have garnered particular interest because of their various transport properties, tunable magnetism, and the strong interplay of magnetic ordering and band topology~\cite{wangColossalMagnetoresistanceMixed2021,wangAnisotropyMagneticTransport2022,berryTypeAntiferromagneticOrder2022,singhSuperexchangeInteractionInsulating2023a}.

Previously, a Weyl semimetal state was identified in EuCd$_2$As$_2$ with the polarized spins of Eu~\cite{huaDiracSemimetalTypeIV2018,maSpinFluctuationInduced2019,wangSinglePairWeyl2019,xuUnconventionalTransverseTransport2021,rahnCouplingMagneticOrder2018}. Additionally, a CMR effect, attributed to strong magnetic fluctuations, was reported in the isostructural EuCd$_2$P$_2$, which rivals the manganates in the CMR magnitude~\cite{wangColossalMagnetoresistanceMixed2021,sunkoSpincarrierCouplingInduced2023,homesOpticalPropertiesCarrier2023}. Furthermore, the properties of the CaAl$_2$Si$_2$-type Eu$M_2X_2$ ($M$ = Zn, Cd; $X$ = P, As, Sb) are extremely sensitive to carrier density in the material. For instance, EuZn$_2$P$_2$ could be transformed from an antiferromagnetic (AFM) insulator to a ferromagnetic (FM) metal through the introduction of Eu vacancies during crystal growth~\cite{Chen2024}. Similarly, the magnetic ground states of other Eu$M_2X_2$ are have been successfully switched from AFM to FM ordering with a low carrier concentration~\cite{Chen2024,joManipulatingMagnetismTopological2020,roychowdhuryAnomalousHallConductivity2023,Chen2024a}. These intriguing findings within Eu$M_2X_2$ family underscore the rich phenomena and highly tunable properties in the magnetic Zintl phases with small band gaps, thereby motivating further exploration of Eu-containing Zintl phases with similar constituents and related structures.

In this work, we present comprehensive characterizations of another Eu-based Zintl phase, Eu$_{11}$Zn$_6$As$_{12}$, which was previously synthesized but whose properties have not been reported~\cite{saparovUndecaeuropiumHexazincDodecaarsenide2010}. Eu$_{11}$Zn$_6$As$_{12}$ is constructed from chains of Eu cations and anionic frameworks composed of corner-sharing ZnAs$_4$ tetrahedra. The $MX_4$ tetrahedron is also the basic structural fragment of Eu$M_2X_2$. Eu$_{11}$Zn$_6$As$_{12}$ crystallizes in a Sr$_{11}$Cd$_6$Sb$_{12}$-type structure and displays complex magnetic behavior resulting from the competing AFM and FM interactions, with strong short-range FM fluctuations around 29 K ($T_\mathrm{F}$), an AFM ordering at 22 K ($T_\mathrm{N}$), and a third transition at 9 K ($T^*$). Moreover, Eu$_{11}$Zn$_6$As$_{12}$ exhibits metallic conductivity with a pronounced resistivity peak around $T_\mathrm{N}$. The resistivity peak is suppressed rapidly in the field, leading to a large negative magnetoresistance (MR) effect near $T_\mathrm{N}$. In addition, a small hysteresis in resistivity below 5 K is also noted, attributed to weakly FM magnetization for the uncompensated Eu spins. These novel features are absent in other Eu-containing 11-6-12 phases, such as Eu$_{11}$Zn$_6$Sb$_{12}$ and Eu$_{11}$Cd$_6$Sb$_{12}$~\cite{saparovSynthesisStructurePhysical2008}. Our findings underscore the surprising differences in the physical properties between these sister compounds, suggesting potential avenues for exploring unusual physical phenomena, such as the pronounced negative MR effect and nonlinear AHE, through full ionic substitution in narrow-gap rare-earth Zintl phases.

\section{Methods}
\subsection{Crystal growth}

High-purity Eu pieces ($99.999\%$), Zn powder ($99.99\%$), and As lumps ($99.99\%$) were used to grow single crystals of Eu$_{11}$Zn$_6$As$_{12}$ in Pb flux ($99.999\%$). The Eu pieces were meticulously cleaned using a cutting plier to remove any oxides prior to use. All reagents and products were handled within an argon-filled glove box to prevent any potential oxidation or hydrolysis.

During an attempt to synthesize Eu$_2$ZnAs$_2$ with the Yb$_2$CdSb$_2$-type structure, we inadvertently obtained Eu$_{11}$Zn$_6$As$_{12}$ single crystals instead. The starting materials of Eu, Zn, As, and Pb, in a molar ratio of 2:1:2:30, were loaded into the alumina crucible with a total weight of 7 g. The crucible was sealed in an evacuated silica tube and heated directly to 1100$^\circ$C over a period of 15 hours, then held at this temperature for 35 hours before being cooled down to 550$^\circ$C at a rate of 3$^\circ$C/h. The silica tube was centrifuged and broken within the glove box. The products of Eu$_{11}$Zn$_6$As$_{12}$ crystals are predominantly lustrous platelets that appear to be consist of transversely arranged, acicular-like single crystals (see Fig.~\ref{F1}(c)). Nonetheless, isolated single crystals were still obtainable for structure determination and measurement purposes.

The Eu$_{11}$Zn$_6$As$_{12}$ samples are air-stable.

\subsection{Structure determination}

Suitable crystals were selected for single-crystal X-ray diffraction (SCXRD) analysis. Data collection was carried out at 150 K using a Bruker D8 Venture diffractometer, which was equipped with an I$\mu$S 3.0 Dual Wavelength system (Mo $K\alpha$ radiation, $\lambda$ = 0.71073 \AA) and an APEX-II CCD detector. The collected frames were reduced and corrected using the Bruker SAINT software suite. The crystal structure was determined using the intrinsic phasing method implemented in the SHELXT structure solution program within the Olex2 environment~\cite{26}. Subsequently, anisotropic refinements were performed with the SHELXL refinement package utilizing the least-squares method~\cite{27}. Finally, the crystal structure was visualized using the VESTA software~\cite{28}. Additionally, no twinning of the crystals was detected for our Eu$_{11}$Zn$_6$As$_{12}$ sample.

\subsection{Magnetization, heat capacity and resistivity measurements}
The zero-field-cooling (ZFC) and field-cooling (FC) direct-current (dc) magnetization measurements of Eu$_{11}$Zn$_6$As$_{12}$ were conducted using a Magnetic Property Measurement System (MPMS 3, Quantum Design). The alternating-current (ac) susceptibility was collected on on a Physical Property Measurement System (PPMS Dynacool, Quantum Design) with an ac measurement system (ACMS II). The temperature sweep rate is 2 K/min for the temperature-dependent magnetization measurements. The heat capacity measurement was also carried out on PPMS Dynacool with a relaxation-time technique. The transport measurements were performed on the TeslatronPT platform (Oxford Instruments) with a 12 T magnet, utilizing a custom-made probe coupled with an SR830 lock-in amplifier and 2400 meter (Stanford Research Systems). The magnetotransport data were measured with the current along the $b$ axis ($\rho_b$) and the field perpendicular to it. High-quality single crystals were carefully selected and polished into the rectangular prism to facilitate resistivity calculations. Transport data were collected with a standard four-probe technique.

\subsection{First-principles calculations}

Density functional theory (DFT) calculations were performed using the Vienna \textit{ab initio} Simulation Package (VASP)~\cite{Kresse1996a}. Electron-ion interactions were described using projector augmented wave (PAW) pseudopotentials~\cite{Blochl1994a}. Due to the failure of density functionals in accurately describing partially filled 4$f$ states, the PAW pseudopotentials were Eu\_2, As, and Zn, as recommended by VASP. The plane-wave cutoff energy was set to 350 eV. The exchange-correlation functional was treated using the Perdew-Burke-Ernzerhof (PBE) parametrization of the generalized gradient approximation (GGA)~\cite{Perdew1996}. A $\Gamma$-centered $1 \times 7 \times 3$ Monkhorst-Pack $k$-mesh was used for Brillouin zone sampling. Both the lattice constants and atomic positions were fully optimized iteratively until the total energy and the Hellmann-Feynman force on each atom converged to $10^{-6}$ eV and 0.05 eV/\AA, respectively.

\section{RESULTS AND DISCUSSION}

\subsection{Crystal structure}

\begin{figure}
	\includegraphics[width=0.48\textwidth]{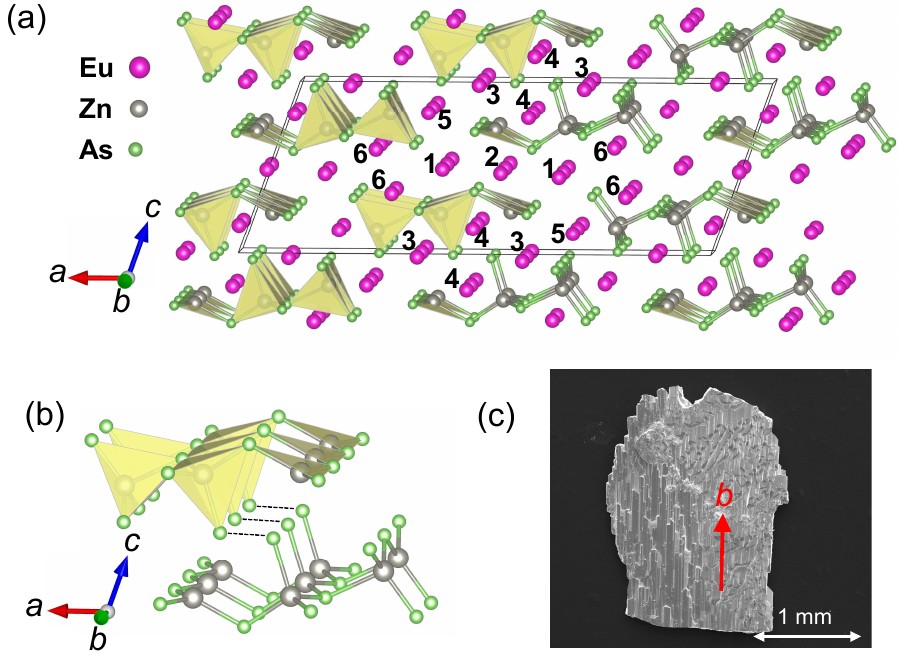}
	\caption{(a) Combined polyhedral and ball-and-stick representations for the crystal structure of Eu$_{11}$Zn$_6$As$_{12}$. The ZnAs$_4$ tetrahedra is highlighted by faint yellow. Eu, Zn and As atoms are shown as pink, gray and green balls, respectively. The Eu sites are labeled with numbers. The structure of Eu$_{11}$Zn$_6$As$_{12}$ is viewed from the crystallographic $c$ axis. Unit cell is outlined. (b) Further structural detail on [Zn$_6$As$_{12}$]$^{22-}$ networks in Eu$_{11}$Zn$_6$As$_{12}$. (c) A micrograph of single crystal of Eu$_{11}$Zn$_6$As$_{12}$.}
	\label{F1}
\end{figure}

Eu$_{11}$Zn$_6$As$_{12}$ crystallizes in the monoclinic Sr$_{11}$Cd$_6$Sb$_{12}$-type structure (space group $C2/m$, No. 12), as shown in Fig.~\ref{F1}(a). The refined structure from SCXRD is listed in Table~\ref{Tab-1}. All cell dimensions of our crystal are 0.2\% smaller than those reported in the literature, owing to the lower measurement temperature~\cite{saparovUndecaeuropiumHexazincDodecaarsenide2010}. Eu$_{11}$Zn$_6$As$_{12}$ consists of Eu cations and covalently bonded polyanionic ribbons, formed through corner-sharing, which can be understood in the following way. Firstly, infinite chains of $_\infty^1[\mathrm{ZnAs}_3]$ are formed by connecting ZnAs$_4$ tetrahedra along the crystallographic $b$ axis. Then two complete $_\infty^1[\mathrm{ZnAs}_3]$ chains and a third incomplete $_\infty^1[\mathrm{ZnAs}_2]$ chain, lacking a corner, integrate to constitute a polyanionic segment designated as $_\infty^1[\mathrm{Zn_3As_6}]$. Two segments are then bridged through a As$_2$ dimer, giving rise to a sub-network of $_\infty^1[\mathrm{Zn_6As_{12}}]^{22-}$, illustrated in Fig.~\ref{F1}(b). And the interstitial spaces are occupied by the Eu cations. Note that three Eu chains (two of Eu1 and one of Eu2), situated at the center of the unit cell, are arranged linearly with uniform spacing owing to rotational symmetry. This configuration may facilitate the formation of FM domains, which will be addressed further in the discussion.

As mentioned in the METHODS section, attempts to synthesize Eu$_2$ZnAs$_2$ were unsuccessful, despite the existence of the Cd analog, Eu$_2$CdAs$_2$~\cite{wangSynthesisCrystalElectronic2011}. The difference in radius between As and the transition metal Zn plays a crucial role in the final phase formation. Based on the existing Zintl phases Eu$_2MX_2$ and Eu$_{11}M_{6}X_{12}$ ($M$ = Cd, Zn; $X$ = P, As, Sb)\cite{wangSynthesisCrystalElectronic2011,saparovUndecaeuropiumHexazincDodecaarsenide2010,saparovSynthesisStructurePhysical2008,wangNewTernaryPhosphides2013}, the trend is clear: the 2-1-2 phase (Yb$_2$CdSb$_2$-type) is favored when the radius of $M$ is significantly larger than that of $X$, whereas comparable radii or a smaller $M$ favor the formation of the 11-6-12 phase.

\begin{table}
	\caption{Crystallographic data and refinement result of Eu$_{11}$Zn$_6$As$_{12}$ at 150 K.}
	\begin{ruledtabular}\label{Tab-1}
		\begin{tabular}{ll}	
			Material  & Eu$_{11}$Zn$_6$As$_{12}$ \\
			\hline
			Crystal system  & monoclinic\\	
			Space group  & $C2/m$ (No. 12)  \\   
			$a$ (\AA)  & 30.2522(18) \\                               
			$b$ (\AA)  &  4.3256(2) \\
			$c$ (\AA)  &  11.7518(5) \\
			$V$ (\AA$^3$)  & 1447.44(13) \\                           
			$Z$  & 2 \\
			$\rho_\mathrm{calc}$ (g/cm$^3$)  &  6.798 \\
			Temperature (K)  & 150 \\
			Radiation  & Mo $K\alpha$ \\
			Reflections collected  & 16150 \\
			Independent reflections  & 1722 \\
			$R\rm_{int}$ & 0.0871 \\
			Goodness-of-fit  & 1.115 \\
			$R_1$\footnote{$R_1=\Sigma||F_o|-|F_c||/\Sigma|F_o|$.}  & 0.0339 \\
			$wR_2$\footnote{$wR_2=[\Sigma w(F_o^2-F_c^2)^2/\Sigma w(F_o^2)^2]^{1/2}$.}   & 0.0714 \\		
		\end{tabular}
	\end{ruledtabular}
\end{table}

\subsection{Magnetic properties}

\begin{figure*}
	\includegraphics[width=0.98\textwidth]{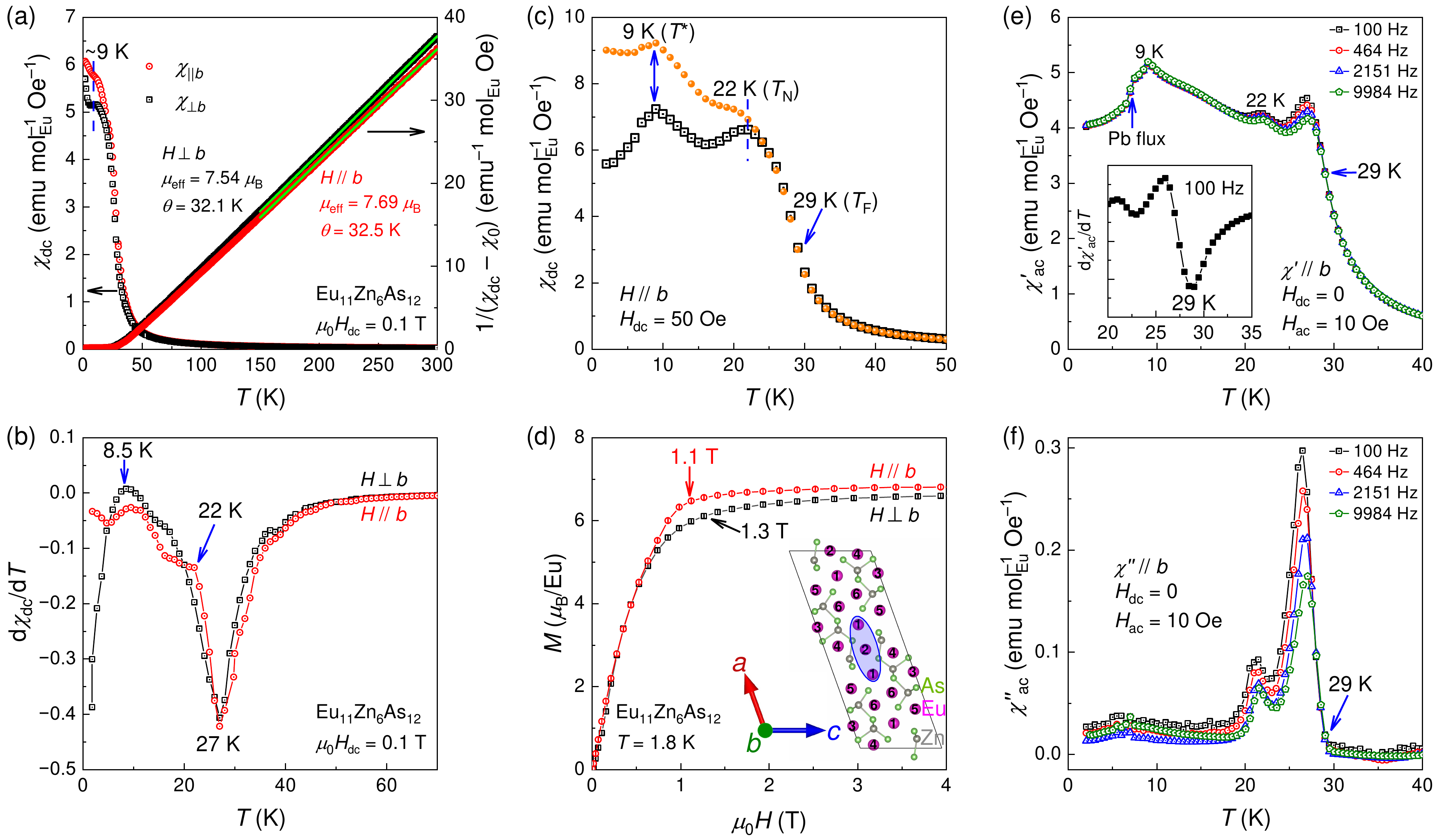}
	\caption{(a) Susceptibility (left axis) of Eu$_{11}$Zn$_6$As$_{12}$ and the Curie-Weiss analysis (right axis). Red circles and black squares represent $\chi(T)$ along and perpendicular to the $b$ axis, respectively. The corresponding Curie-Weiss fit is plotted as the green line on the reciprocal of susceptibility. (b) $d\chi/dT$ as a function of temperature to show the magnetic transitions of Eu$_{11}$Zn$_6$As$_{12}$. (c) The bifurcation of ZFC (black) and FC (orange) data collected under a field of 5 mT along the $b$ axis. (d) Magnetization of Eu$_{11}$Zn$_6$As$_{12}$ as a function of fields along (red circles) and perpendicular to  (black squares) the $b$ axis. Inset shows a projection of Eu$_{11}$Zn$_6$As$_{12}$ structure along [010]-direction and highlight the Eu1 and Eu2 sites. (e) The in-phase component of ac magnetic susceptibility $\chi^\prime(T)$ at different frequencies. The ac driving amplitude was 10 Oe and the dc magnetic field is zero. Inset displays the $d\chi^\prime/dT$ as a function of $T$. (f) The out-of-phase component of ac magnetic susceptibility $\chi^{\prime\prime}(T)$ at different frequencies.}
	
\label{F2}
\end{figure*}

Eu$_{11}$Zn$_6$As$_{12}$ exhibits a complicated magnetic behavior, which is summarized in Fig.~\ref{F2}. Panel (a) presents the dc susceptibility and inverse susceptibility of Eu$_{11}$Zn$_6$As$_{12}$ under $\mu_0H_\mathrm{dc} = 0.1$ T. The anisotropy of $\chi_{\parallel b}(T)$ and $\chi_{\perp b}(T)$ is small. Notably, both $\chi_{\parallel b}(T)$ and $\chi_{\perp b}(T)$ increase obviously below 50 K. For $H\parallel b$, $\chi_{\parallel b}(T)$ keeps increasing at low temperatures and a kink at about 9 K is observed. While for $H\perp b$, $\chi_{\perp b}(T)$ deviates conspicuously from $\chi_{\parallel b}(T)$ around 18 K and becomes flat at 11 K, exhibiting a plateau-like behavior. Below 5 K, $\chi_{\perp b}(T)$ continues to increase and almost catches up with $\chi_{\parallel b}(T)$ at 1.8 K. By fitting the $\chi(T)$ data from 150 to 300 K to the Curie-Weiss law $\chi(T) - \chi_\mathrm{0} = C/(T - \theta)$, a Weiss temperature of $\theta\approx32$ K and an effective moment of $\mu_\mathrm{eff}\approx7.6$ $\mu_\mathrm{B}$/Eu are obtained. The positive $\theta$ is indicative of strong FM interaction in Eu$_{11}$Zn$_6$As$_{12}$. $\mu_\mathrm{eff}$ of Eu$_{11}$Zn$_6$As$_{12}$ is a bit lower than the theoretical value of 7.94 $\mu_\mathrm{B}$ for Eu$^{2+}$, may suggest the possibility of a slightly mixed valence of Eu$^{2+}$ and Eu$^{3+}$ or a small proportion of Eu vacancies. The comparatively low $\mu_\mathrm{eff}$ is consistent with the saturation magnetizations ($M_\mathrm{sat}$) shown in Fig.~\ref{F2}(d), which are also slightly below the theoretical $M_\mathrm{sat}$ of 7 $\mu_\mathrm{B}$.

The temperature-dependent derivatives of susceptibility ($d\chi_\mathrm{dc}/dT$) are plotted in Fig.~\ref{F2}(b) to exhibit the features related to the magnetic transitions. A negative peak is situated at 27 K, corresponding to the rise below 50 K in Figs.~\ref{F2}(a) and \ref{F2}(c). Generally, the rapid rise in susceptibility indicates the presence of FM correlations, which is confirmed by a small bifurcation of the ZFC and FC data under a small field of 50 Oe in Fig.~\ref{F2}(c), as well as the large out-of-phase contribution of ac susceptibility ($\chi^{\prime\prime}_{\mathrm{ac}}\sim7\%\chi^{\prime}_{\mathrm{ac}}$) below 29 K, as shown in Figs.~\ref{F2}(e) and \ref{F2}(f). However, it is more appropriate to regard the transition around 27 K as a sign of short-range FM fluctuations rather than the establish of long-range FM order based on the following observations: (i) The signature of transition around 27 K is absent in the temperature-dependent specific heat ($C_p$) in Fig.~\ref{F5}(a), although a weak rise in $C_p/T$ from 28 K is seen in Fig.~\ref{F5}(c); (ii) The magnetic anisotropy of susceptibility in Figs.~\ref{F2}(a) and \ref{F2}(c) is slight, unlike a typical ferromagnet; (iii) The splitting of ZFC and FC data in Fig.~\ref{F2}(c) is rather small; (iv) The rise of susceptibility around 27 K is unremarkable, unlike other reported Eu-based ferromagnets~\cite{Chen2024,Chen2024a,joManipulatingMagnetismTopological2020,roychowdhuryAnomalousHallConductivity2023,devlinEu11Zn2018}. One may note that the $\chi^{\prime\prime}_{\mathrm{ac}}$ peak for the fluctuations slightly shifts to higher temperature by about 0.6 K when the frequency increases from 100 Hz to 9984 Hz, which does not indicate a glassy behavior, since weak frequency dependence can also be observed when FM phase appears~\cite{APPA}. Moreover, we define the temperature of the $d\chi^\prime_\mathrm{ac}/dT$ minimum or the $\chi^{\prime\prime}_\mathrm{ac}$ onset, i.e. 29 K, as the characteristic temperature ($T_\mathrm{F}$) of FM fluctuations.

The first long-range order occurs around 22 K, as evidenced by the plateau in Fig.~\ref{F2}(c) and the small peaks of $\chi_{\mathrm{ac}}$ in Figs.~\ref{F2}(e) and \ref{F2}(f). This transition is not conspicuous for $\chi_{\mathrm{dc}}(T)$ in Fig.~\ref{F2}(a); only a small kink is visible from $d\chi_{\mathrm{dc}}/dT$ in Fig.~\ref{F2}(b). Nonetheless, this transition should be considered as a long-range order given the evident changes of heat capacity and resistivity close to the corresponding temperature, which will be discussed later. The long-range order is likely AFM since $\chi_\mathrm{dc}(T)$ does not show a sudden increase below 22 K. It is worth noting that the typical decline in $\chi_\mathrm{dc}(T)$ resulting from the AFM transition is subtle, not only due to the strong background from FM fluctuations mentioned earlier, but also because of the low-dimension magnetic correlations arising from the chain-like structure of Eu$_{11}$Zn$_6$As$_{12}$. It is common to observe the absence of a sharp transition peak for quasi-one-dimensional or quasi-two-dimensional antiferromagnets~\cite{T_N_2,BaNaCuVO_1D,wangAbsenceSuperconductivityElectrondoped2023}. Additionally, the presence of a $\chi^{\prime\prime}_{\mathrm{ac}}$ peak around 22 K, which is typically absent in AFM materials, can be attributed to the background FM fluctuations and the fact that only the spins at partial Eu sites are aligned antiparallelly. Hence, the AFM transition temperature ($T_\mathrm{N}$) of Eu$_{11}$Zn$_6$As$_{12}$ is 22 K according to the susceptibility criteria.

A clear transition peak at 9 K is observed for both dc susceptibility (ZFC and FC) in Fig.~\ref{F2}(c) and in-phase ac susceptibility $\chi^{\prime}_{ac}$ in Fig.~\ref{F2}(e). Considering the evident heat capacity jump at 8.5 K in Fig.~\ref{F5}(a), we believe this is a bulk transition resulting from the spin reorientation or the ordering of the remaining portion of Eu sites. Moreover, this transition seems to be AFM since the response of $\chi^{\prime\prime}_{\mathrm{ac}}$ around 9 K almost vanishes. The intricate magnetic behavior of Eu$_{11}$Zn$_6$As$_{12}$ could be ascribed to its complex crystal structure. Given the predominant FM interaction indicated by the positive Weiss temperature, it is plausible to assume FM coupling along the Eu chains for the higher susceptibility and magnetization along the $b$ axis. Moreover, for Eu$_{11}$Zn$_6$As$_{12}$, three Eu atoms (Eu1-Eu2-Eu1) are arranged linearly (bond angle 180$^\circ$) in [010] plane with a distance ($d_\mathrm{Eu1-Eu2}$) of 3.736 \AA, as highlighted in the inset of Fig.~\ref{F2}(d). The distance is much shorter than the Eu-Eu distance (4.326 \AA) along the Eu chain ($b$ axis) and the in-plane Eu-Eu distances of many other Eu-based layered materials ($4\sim5$ \AA)~\cite{rahnCouplingMagneticOrder2018,berryTypeAntiferromagneticOrder2022}, which may favor the FM coupling for the enhanced exchange interaction and dipolar interaction~\cite{johnstonMagneticDipoleInteractions2016,Pakhira2020}. Therefore, it is reasonable to suppose a FM coupling between Eu1 and Eu2 chains of Eu$_{11}$Zn$_6$As$_{12}$. The interchain FM coupling may exist not only between Eu1 and Eu2, but also between chains of Eu1-Eu5 ($d_\mathrm{Eu1-Eu5}=3.711$ \AA, Eu5-Eu1-Eu2 bond angle 107.6$^\circ$). The ferromagnetically coupled Eu chains are coupled through the AFM interactions mediated by the network of $_\infty^1[\mathrm{Zn_6As_{12}}]$. However, due to the crooked anionic segments, perfect compensation of ferromagnetically coupled units may be difficult, resulting in a small net magnetization that leads to the splitting of the ZFC and FC data below $T_\mathrm{F}$, as well as the resistivity hysteresis near zero temperature in Fig.~\ref{F4}. Moreover, the spins of Eu may order progressively due to the varied exchange interactions between the multiple magnetic Wyckoff sites. Hence, it is also possible that the three magnetic transitions occur on different Eu sites at successive temperatures.

By comparison, Eu$_{11}$Zn$_6$Sb$_{12}$ and Eu$_{11}$Cd$_6$Sb$_{12}$, the sibling compounds of Eu$_{11}$Zn$_6$As$_{12}$, are nonmagnetic or clearly AFM, and their Weiss temperature are below 0 K~\cite{saparovSynthesisStructurePhysical2008}. The sharp contrast is attributed to the obviously longer Eu-Eu distances in the antimonides, including the intra- ($\gtrsim$ 4.5 \AA) and interchain ($\gtrsim$ 3.9 \AA) distances. Of course, further experiments, such as neutron scattering, are necessary to determine the magnetic structure of Eu$_{11}$Zn$_6$As$_{12}$.

\subsection{Heat capacity}

\begin{figure}
	\includegraphics[width=0.45\textwidth]{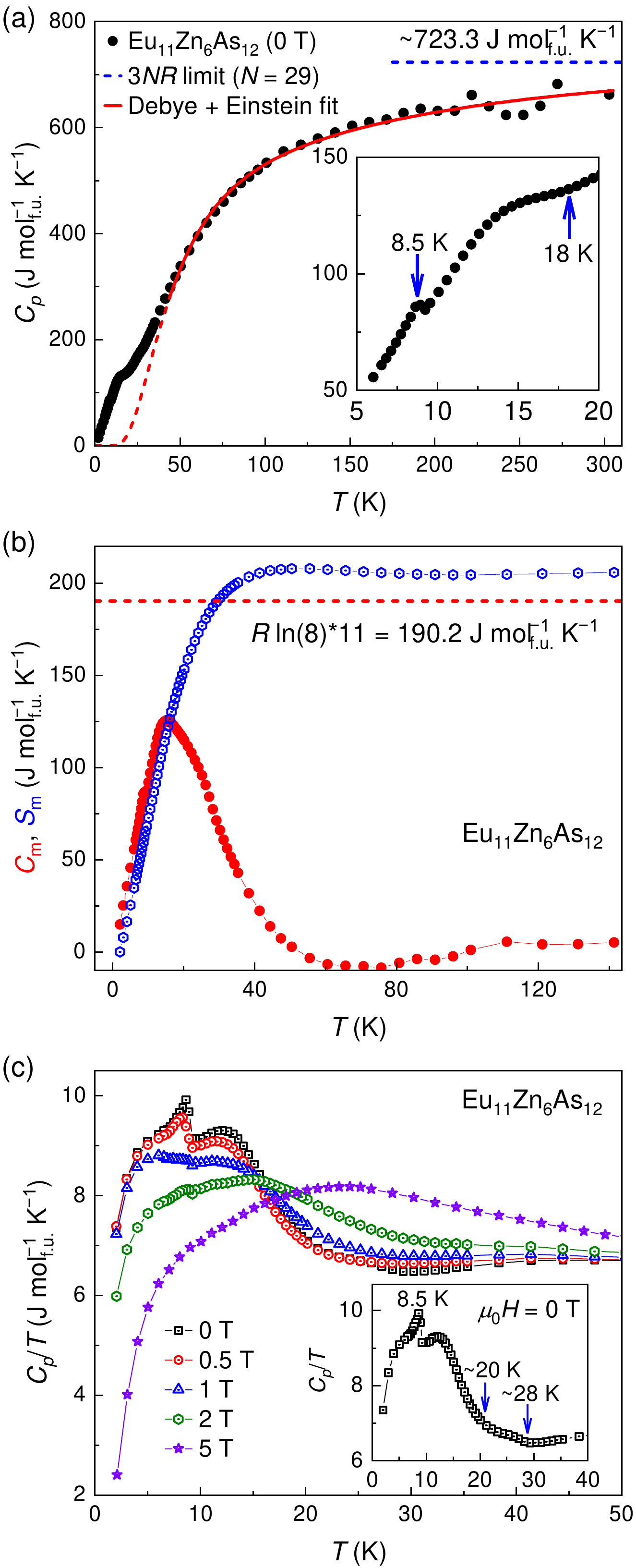}
	\caption{(a) Temperature dependence of zero-field heat capacity, $C_p(T)$, along with a fit from 45 to 300 K by a combination of the Debye and Einstein models (red curve). Inset zooms in the $C_p(T)$ data below 20 K to highlight the transitions. (b) Magnetic heat capacity $C_\mathrm{m}(T)$ (red dots) and entropy $S_\mathrm{m}(T)$ (blue circles) calculated from $C_\mathrm{m}(T)$, plotted as a function of temperature. The red dashed line marks the magnetic entropy expected for Eu$^{2+}$ ions. (c) $C_p/T$ vs $T$ under various magnetic fields. Inset plots zero-field $C_p/T$ vs $T$ individually to show the transitions.}
	\label{F5}
\end{figure}

To confirm the magnetic transitions of Eu$_{11}$Zn$_6$As$_{12}$ revealed by the magnetization data, we measured the specific heat ($C_p$) and present the result in Fig.~\ref{F5}. As shown in Fig.~\ref{F5}(a), the $C_p$ value at room temperature is smaller than the expected Dulong-Petit limit of 723.3 J mol$^{-1}_\mathrm{f.u.}$ K$^{-1}$ ($N$ = 29) for Eu$_{11}$Zn$_6$As$_{12}$. In the lower right inset, a broad hump attributed to the AFM transition is observed below 18 K, corresponding to the resistivity peak at this temperature and slightly lower than the $T_\mathrm{N}$ value determined by the susceptibility. The $C_p$ of Eu$_{11}$Zn$_6$As$_{12}$ around $T_\mathrm{N}$ does not exhibit a $\lambda$-type peak, consistent with its structure of infinite chains, since the $\lambda$-type peak is typically absent in materials with low-dimensional AFM correlations~\cite{T_N,T_N_2}. With decreasing temperature, a small but clear anomaly occurs at 8.5 K, corresponding to the second bulk transition ($T^*$) discussed earlier. We fit the zero-field $C_p$ data from 45 to 300 K by a combination of the Debye and Einstein models according to
\begin{equation}
	C_p(T)=\alpha C_\mathrm{D}(T)+(1-\alpha)C_\mathrm{E}(T),
\end{equation}
where
\begin{eqnarray}
	C_\mathrm{D}(T)=9NR(\frac{T}{\mathrm{\Theta_D}})^3\int^{\mathrm{\Theta_D}/T}_0\frac{x^4e^x}{(e^x-1)^2}dx,\\
	C_\mathrm{E}(T)=3NR(\frac{\mathrm{\Theta_E}}{T})^2\frac{e^{\mathrm{\Theta_D}/T}}{(e^{\mathrm{\Theta_D}/T}-1)^2}.
\end{eqnarray}
Here, $C_\mathrm{D}(T)$ and $C_\mathrm{E}(T)$ represent the lattice contributions from the Debye and Einstein models, respectively. $\mathrm{\Theta_D}$ and $\mathrm{\Theta_E}$ are the Debye temperature and Einstein temperature, and $\alpha$ determines the relative contributions of the Debye and Einstein components. The fitted parameters are $\mathrm{\Theta_D}=1025.1$ K, $\mathrm{\Theta_E} = 140.6$ K, and $\alpha = 0.15$. 

With the fitted data, the magnetic contribution to the heat capacity ($C_\mathrm{m}$) was obtained by subtracting the lattice contribution, and the magnetic entropy ($S_\mathrm{m}$) was calculated by the integral
\begin{equation}
	S_\mathrm{m}(T) = \int^T_0 \dfrac{C_\mathrm{m}(T)}{T}dT,
\end{equation}
which are plotted in the Fig.~\ref{F5}(b). The resultant $S_\mathrm{m}$ is about 205 J mol$^{-1}_\mathrm{f.u.}$ K$^{-1}$, slightly exceeding the expected value of $S_\mathrm{m}=11R\ln(2S+1)$ with $S=7/2$. This excess may be attributed to an overestimation of the Einstein component due to the data fluctuation above 200 K. Additionally, the evident contributions to $C_\mathrm{m}(T)$ and $S_\mathrm{m}(T)$ above $T_\mathrm{N}$ arise from short-range magnetic interactions.

Figure~\ref{F5}(c) shows the $C_p/T$ data under several magnetic fields. The sharp peak at 8.5 K confirms the bulk transition at $T^*$. The broad shoulder below $T^*$ is commonly observed in $S = 7/2$ magnets~\cite{Pakhira2020,johnston2015UnifiedMolecularField}. Furthermore, as shown in the inset, two successive rises are observed starting around 28 K and 20 K, contributed by the magnetic fluctuations around $T_\mathrm{F}$ and AFM ordering at $T_\mathrm{N}$, respectively. Upon applying the fields, the peak at $T^*$ is suppressed and disappears substantially above 2 T. Moreover, the phase transitions at $T_\mathrm{N}$ and $T_\mathrm{F}$ are severely broaden by the fields, resulting in a weight shift of heat capacity from lower to higher temperature.

\subsection{Magnetotransport}

\begin{figure*}
	\includegraphics[width=0.9\textwidth]{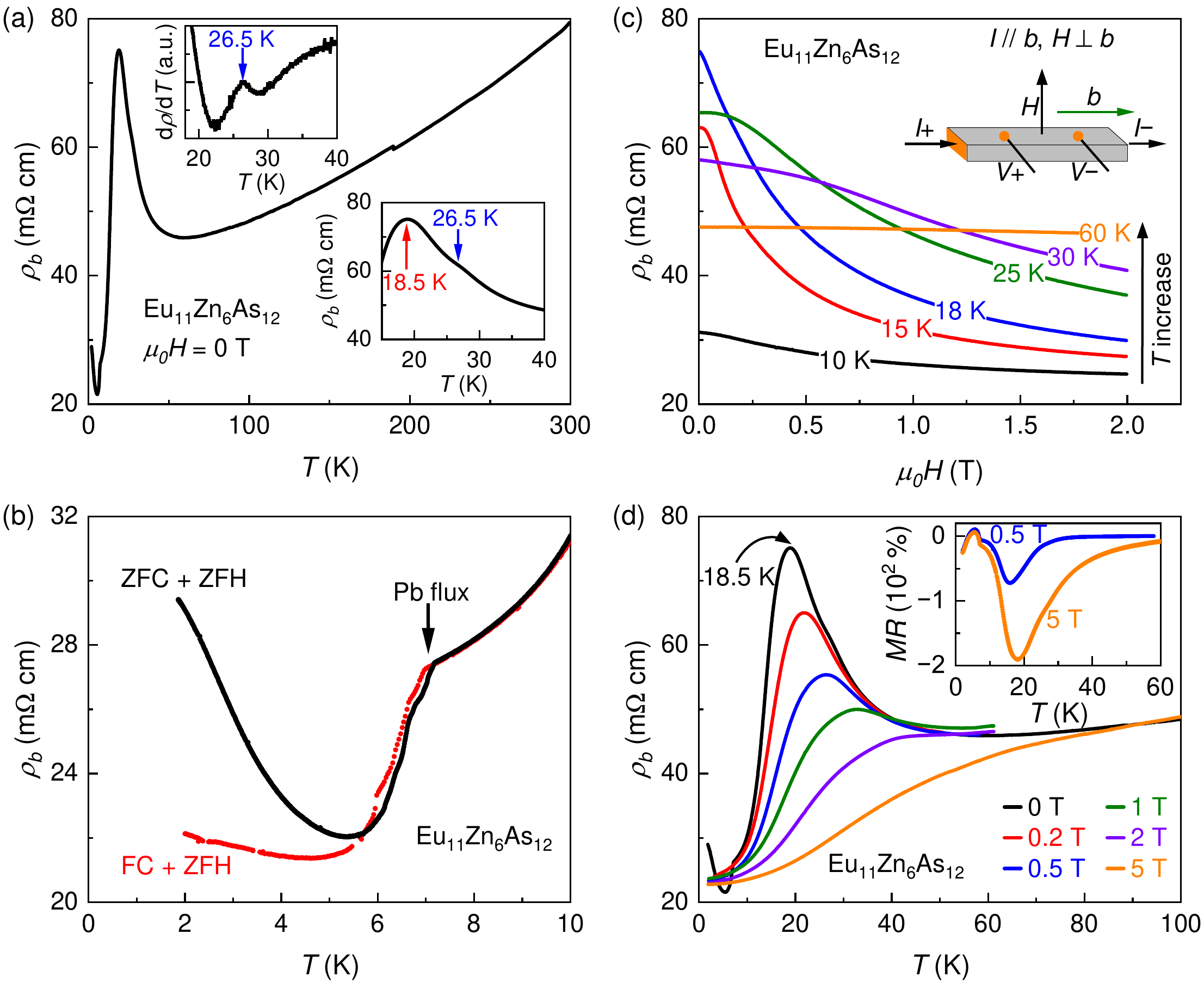}
	\caption{(a) Temperature-dependent electrical resistivity of Eu$_{11}$Zn$_6$As$_{12}$ along the $b$ axis. The upper left inset shows the derivative of $\rho_b(T)$, while the lower right inset zooms in $\rho_b(T)$ to show the transitions. (b) The hysteresis of $\rho_b(T)$. The black dots represent the data measured in zero-field-heating (ZFH) process after cooling to 2 K without the field (ZFC), and the red dots represent the data measured in ZFH process after cooling to 2 K under 5 T (FC). (c) Field-dependent $\rho_b$ at various temperatures from 10 to 60 K. Inset illustrates the electrode configuration for the resistivity measurement. (d) Temperature-dependent $\rho_b$ with applied magnetic field of 0, 0.2, 0.5, 1, 2 and 5 T. Inset shows the MR as a function of temperature at 0.5 and 5 T.}
	\label{F4}
\end{figure*}

Eu$_{11}$Zn$_6$As$_{12}$ exhibits distinct electrical transport properties from the isostructural Eu$_{11}$Cd$_6$Sb$_{12}$ and Eu$_{11}$Zn$_6$Sb$_{12}$, which were reported to show simple metallic behavior~\cite{saparovSynthesisStructurePhysical2008}. The electrical resistivity along the $b$ axis ($\rho_b(T)$) of Eu$_{11}$Zn$_6$As$_{12}$ under zero field is presented in Fig.~\ref{F4}(a). It shows a mild decrease with decreasing temperature above 50 K and reveals a large peak at 18.5 K, slightly lower than $T_\mathrm{N}$ based on susceptibility criteria. Below the transition, $\rho_b(T)$ drops rapidly for the reduced magnetic scattering. $\rho_b(T)$ is higher than 20 m$\Omega$ cm across the entire temperature range, significantly above the resistivity of Eu$_{11}$Cd$_6$Sb$_{12}$~\cite{saparovSynthesisStructurePhysical2008}. This is attributed to the less orbital hybridization in Eu$_{11}$Zn$_6$As$_{12}$ due to the smaller spatial extension of the Zn and As orbitals. The magnified view of $\rho_b(T)$ between 15 and 40 K is shown in the lower right inset. A weak hump around 26.5 K is seen, consistent with the temperature range of susceptibility rise in Figs.~\ref{F2}(a,c). To highlight the resistivity anomaly, we plot the temperature derivative of $\rho_b(T)$, ${d\rho}/{dT}$, in the upper left inset. An explicit peak is observed at 26.5 K, consistent with the characteristic temperature of FM fluctuations indicated by the susceptibility.

The zero-field resistivity of Eu$_{11}$Zn$_6$As$_{12}$ at low temperatures is affected by the remanent magnetization, i.e., depends on the processing history in the field, manifested in the Fig.~\ref{F4}(b). Both the black and red dots were measured via zero-field-heating (ZFH) process, and the distinction is that the red data was collected following cooling the sample to 2 K in a field of 5 T , whereas the black data was collected without subjecting the sample to any magnetic field treatment. Both ZFC and FC data drop at 7 K, resulting from a small Pb-flux residue on the rugged surface. The evident upturn in the ZFC resistivity is attributed to the magnetic scattering arising from the uncompensated Eu spins. Conversely, the FC data exhibits no substantial increase, as the spins are aligned in the applied magnetic field. The bifurcation between ZFC and FC data validates the existence of the net FM component under low temperatures.

By applying the magnetic field, Eu$_{11}$Zn$_6$As$_{12}$ shows a noticeable negative MR effect, as presented in Figs.~\ref{F4}(c) and~\ref{F4}(d). $\rho_b(H)$ was measured with the field parallel to the $ac$ plane ($H\perp b$). At 60 K, no conspicuous MR effect is observed below 2 T. With the decreasing temperature, the negative MR is enhanced by the strong magnetic fluctuations above the AFM order, and reaches a maximum at $T_\mathrm{N}$ of 18 K. At 18 K, $\rho_b$ declines by 60\% under the field of 2 T. Below $T_\mathrm{N}$, the MR effect is diminished gradually for the reduction of spin-dependent scattering. The temperature-dependent $\rho_b$ at various fields are also depicted in Fig.~\ref{F4}(d). Note that the tail of $\rho_b(0\ T)$ at low temperatures is absent for the curves with field, because both the superconductivity of Pb and magnetic scattering are suppressed by the magnetic field. The resistivity peak at 18.5 K is suppressed rapidly in the field, resulting in a maximum MR of $-190\%$ at 5 T, with the definition that $\mathrm{MR} = 100\%\times[\rho(H) - \rho(0)]/\rho(H)$. The calculated MR at 0.5 T and 5 T are plotted as a function of temperature in the inset of Fig.~\ref{F4}(d). Note that the small positive MR at low temperatures results from the suppression of the superconductivity of Pb, not the intrinsic behavior of Eu$_{11}$Zn$_6$As$_{12}$.

\subsection{Hall effect}

\begin{figure}
	\includegraphics[width=0.48\textwidth]{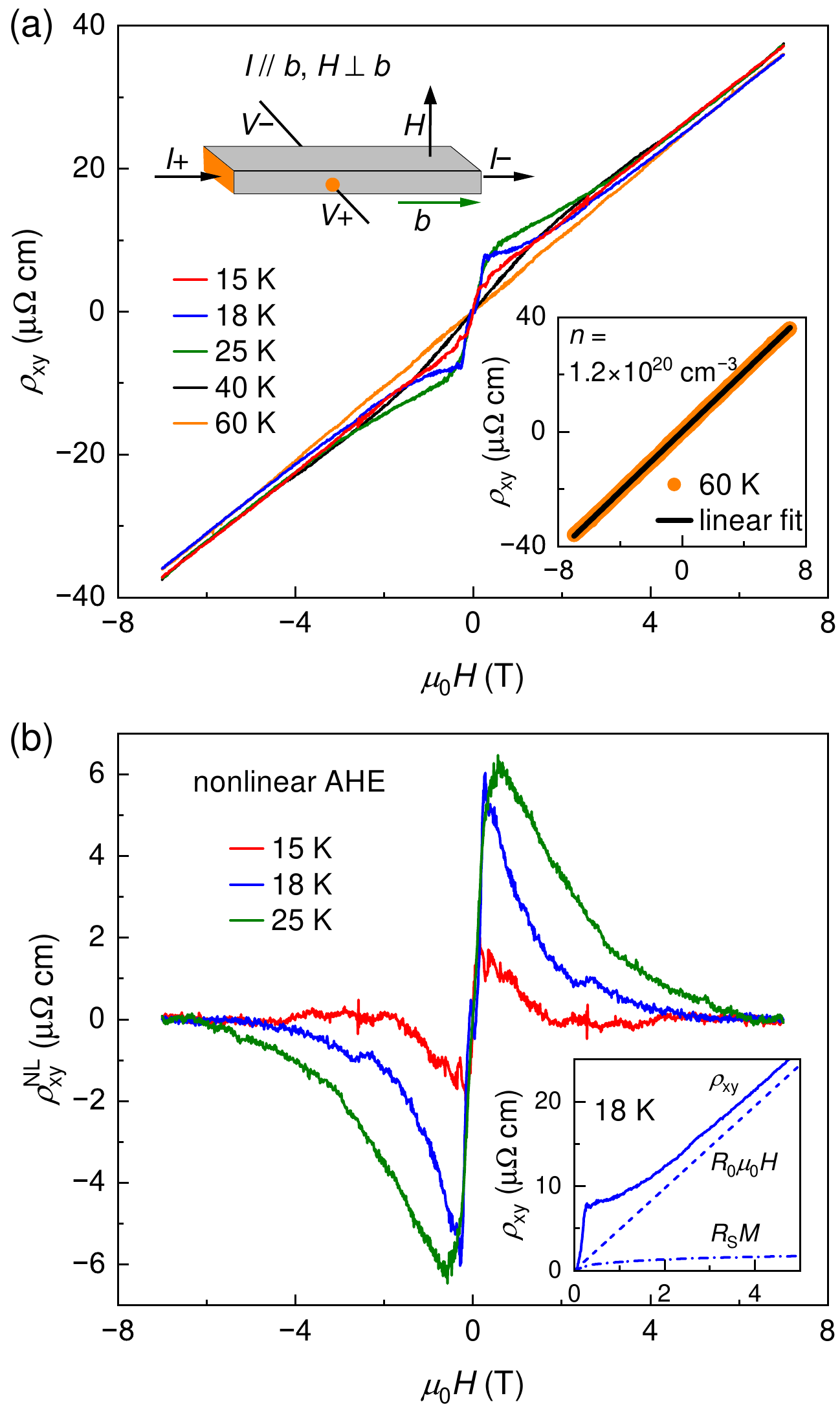}
	\caption{(a) Magnetic field dependence of the Hall resistivity of Eu$_{11}$Zn$_6$As$_{12}$ at several temperatures. The fields are applied perpendicular to the $b$ axis while the current is along the $b$ axis, as illustrated in the upper left schematic. Lower right inset shows a linear fit for the Hall resistivity at 60 K. (b) Extracted nonlinear AHE contribution from Hall resistivity at 15 K, 18 K, and 25 K. Inset shows the decomposition of $\rho_{xy}(H)$ at 18 K, where the total $\rho_{xy}$, OHE contribution, and conventional AHE contribution are plotted as solid, dashed, dash-dotted lines, respectively.}
	\label{F6}
\end{figure}

The field dependence of the Hall resistivity ($\rho_{xy}$) for Eu$_{11}$Zn$_6$As$_{12}$ is shown in Fig.~\ref{F6}(a), with $I\parallel b$ and $H\perp b$. At $T$ = 60 K, $\rho_{xy}$ depends on the field linearly with a positive slope of 5.2 $\mu\Omega$ cm T$^{-1}$, indicating that $\rho_{xy}$ is dominated by the ordinary Hall effect (OHE) when temperature is well above $T_\mathrm{N}$. The carriers are identified as hole type with the density of $n = 1.2\times 10^{20}$ cm$^{-3}$ based on the slope of $\rho_{xy}(H)$. Under a high field ($>4$ T), all $\rho_{xy}(H)$ curves follow a slope similar to the curve at 60 K, suggesting an almost constant carrier density in Eu$_{11}$Zn$_6$As$_{12}$. As the temperature gets closer to 18 K, the curves become more and more nonlinear, which is contributed by the AHE. At 18 K, a small peak is observed at approximately 0.26 T, suggesting the presence of a nonlinear component in addition to the conventional AHE contribution that is proportional to the magnetization. 

We analyzed the nonlinear AHE for the data at 15, 18, and 25 K, as shown in Fig.~\ref{F6}(b). The total Hall resistivity can be expressed as a sum of three parts, $\rho_{xy} = R_0\mu_0H + R_\mathrm{S}M + \rho^{\mathrm{NL}}_{xy}$, where $R_0\mu_0H$ and $R_\mathrm{S}M$ represent the OHE and conventional AHE contributions, respectively. The third term, $\rho^{\mathrm{NL}}_{xy}$, is not proportional to either $\mu_0H$ or $M$ and represents the nonlinear AHE. The magnetization at these temperatures varies mildly at high fields and can be taken as a constant value. The $M(H)$ curves at related temperatures are presented in Fig.~\ref{FA} in the Appendix. Therefore, $R_0\mu_0H + R_\mathrm{S}M$ is only dependent on $\mu_0H$ at high fields. We fit the total Hall resistivity $\rho_{xy}(H)$ from 6 T to 7 T to the equation $R_0\mu_0H + R_\mathrm{S}M$, where $R_0$ is the resulting slope, and $R_\mathrm{S}$ could be obtained from the intercept $R_\mathrm{S}M$. Thus, the nonlinear AHE could be obtained by subtracting OHE and conventional AHE. The decomposition of $\rho_{xy}$ at 18 K is presented in the inset of Fig.~\ref{F6}(b). The extracted $\rho^{\mathrm{NL}}_{xy}$ strongly depends on the temperature and field, exhibiting a clear peak for the $\rho^{\mathrm{NL}}_{xy}(H)$ curves. For the 25 K curve, $\rho^{\mathrm{NL}}_{xy}$ achieves its maximum amplitude at 0.6 T, which constitutes 62\% of the total Hall resistivity ($\rho^{\mathrm{NL}}_{xy}/\rho_{xy}=0.62$). Similar phenomena of large nonlinear AHE contributions have also been reported in other Eu-based Zintl compounds, specifically Eu$M_2X_2$ ($M$ = Zn, Cd; $X$ = P, As, Sb)~\cite{cao2022GiantNonlinearAnomalous,wangAnisotropyMagneticTransport2022,Singh2024}, which are believed to originate from nontrivial band topology or nonzero spin chirality within the domain walls~\cite{xuUnconventionalTransverseTransport2021,Yi2023,Singh2024}. Given the strong FM fluctuations indicated by the magnetic data in Fig.~\ref{F2}, it is not surprising that potentially existing noncollinear spin configurations between the domains host local magnetic chirality and drive the nonzero Berry curvature. The nonlinear AHE contribution diminishes gradually below $T_\mathrm{N}$, as demonstrated by weak $\rho^{\mathrm{NL}}_{xy}$ contribution at 15 K. 

If the hole carriers in Eu$_{11}$Zn$_6$As$_{12}$ are considered to arise from slight Eu defects, the density of Eu vacancies is estimated to be 0.087 per formula unit (or 0.174 hole/f.u.) given $n = 1.2\times 10^{20}$ cm$^{-3}$. That is to say, the proportion of Eu vacancies to all Eu sites is 0.8\%. It is common to see large $p$-type carrier concentrations in Zintl phases for the cation defects, such as the widely studied thermoelectric materials $A$Zn$_2$Sb$_2$ ($A$ = Ca, Sr, Eu, Yb)~\cite{Pomrehn2014,Toberer2010}. Actually, the formation of cation vacancies is energetically favorable in the cation-rich Zintl phases, such as the 11-6-12 phase~\cite{goraiSimpleChemicalGuide2019}. Compared to Eu$_{11}$Zn$_6$Sb$_{12}$ and Eu$_{11}$Cd$_6$Sb$_{12}$, the smaller unit cell of Eu$_{11}$Zn$_6$As$_{12}$ signifies the higher concentration of Eu, implying a higher possibility of Eu vacancies. In addition, recent studies on Eu$M_2X_2$ ($M$ = Cd, Zn; $X$ = P, As) have revealed that FM interaction between Eu layers could be induced or enhanced by the hole carriers~\cite{Chen2024}. Hence, the carriers in Eu$_{11}$Zn$_6$As$_{12}$ may not only account for the metallic behavior, but also be responsible for its enhanced FM coupling.

The origin of carriers in Eu$_{11}$Zn$_6$As$_{12}$ can also be explained in light of the mixed valence of Eu$^{2+}$ and Eu$^{3+}$. A previous study comprehensively discussed the complex chemical bonding and charge balance of the isostructural Zintl compounds, Sr$_{11}$Cd$_6$Sb$_{12}$ and Ba$_{11}$Cd$_6$Sb$_{12}$, affirming that these materials conform to the Zintl concept~\cite{Xia2008}. However, this analysis may not be applicable to rule out the mixed valence in Eu$_{11}$Zn$_6$As$_{12}$. For example, EuNi$_2$P$_2$ is a mixed-valent compound~\cite{EuNi2P2_PRL}, distinct from its siblings with divalent alkaline-earth elements, SrNi$_2$P$_2$ and BaNi$_2$P$_2$~\cite{SrNi2P2_PRB,Mine2008}. The combination of 9 Eu$^{2+}$ and 2 Eu$^{3+}$ could generate a carrier concentration of 0.18 hole/f.u., which is excellently consistent with the value resulting from the Hall effect. Nevertheless, considering that the discrepancies between the experimental and theoretical values of $\mu_\mathrm{eff}$ and $M_\mathrm{sat}$ are not conspicuous, some electrons may transfer from As 4$p$ orbitals to Eu 4$f$ orbitals, not changing the hole density but reducing the ratio of Eu$^{3+}$.

Both assumptions, the presence of Eu vacancies in the lattice and the mixed valence of Eu$^{2+}$ and Eu$^{3+}$, make sense to elucidate the metallic behavior in Eu$_{11}$Zn$_6$As$_{12}$. However, further experiments are necessary to verify these assumptions.

\subsection{DFT calculations}

\begin{figure*}
	\includegraphics[width=0.98\textwidth]{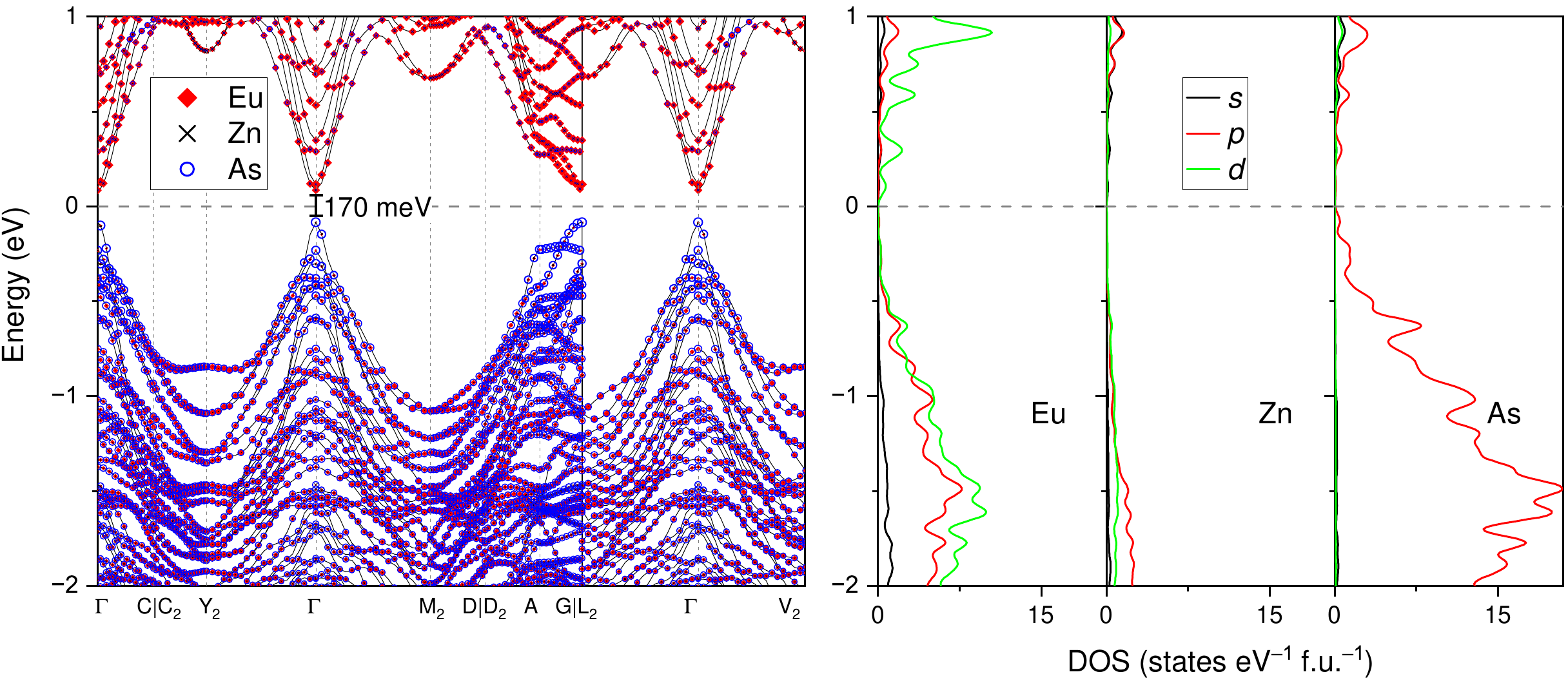}
	\caption{Band structure and DOS of Eu$_{11}$Zn$_6$As$_{12}$, calculated by treating the Eu 4$f$ electrons as core states.}
	\label{F7}
\end{figure*}

To gain deeper insight into the electronic properties, we conducted DFT calculations for Eu$_{11}$Zn$_6$As$_{12}$. However, attempts to treat the half-filled 4$f$ electrons of Eu$^{2+}$ ($4f^7$) as valence electrons failed to converge. Generally, partially filled 4$f$ states cannot be accurately described by current density functionals due to self-interaction errors, except in the case of a half-filled 4$f$ shell, as implemented in Refs.~\cite{Chen2024a,Yang2024}. In contrast, a smooth self-interaction process was achieved when we placed the 4$f$ electrons in the core. Based on this observation, we speculate that the valence of Eu may not be entirely divalent, since Eu$^{2+}$ should be adequately handled by the DFT calculations.

Since the failure to converge when considering the 4$f$ states, we utilized the PAW pseudopotential Eu\_2 to address this challenge, in which the spins of the 4$f$ electrons are omitted. The resulting cell dimensions are: $a = 30.029$ \AA, $b = 4.312$ \AA, and $c = 11.628$ \AA, which are slightly smaller than the experimental values. The band structure and density of states (DOS) are presented in Fig.~\ref{F7}. Although our calculations, without considering magnetism, may not fully reflect the actual electronic band structure of Eu$_{11}$Zn$_6$As$_{12}$, some insights can still be deduced. Firstly, the energy gap of Eu$_{11}$Zn$_6$As$_{12}$ is likely to be narrow and sensitive to the spin arrangement, explaining the reduced resistivity when a magnetic field is applied~\cite{Cuono2023,EuZn2P22023}. Secondly, the 4$p$ orbitals of As dominate the states at the top of the valence band, due to the absence of Eu 4$f$ orbitals. Hybridization of As 4$p$ and Eu 4$f$ orbitals is common in Eu-based pnictides~\cite{Chen2024a,wangSinglePairWeyl2019}, making it reasonable to assume electron transfer between As 4$p$ and Eu 4$f$ orbitals, as discussed previously.

\section{Concluding Remarks}

To summarize, the Eu-containing Zintl phase, Eu$_{11}$Zn$_6$As$_{12}$, was successfully synthesized and carefully characterized. Eu$_{11}$Zn$_6$As$_{12}$ exhibits unusual magnetic and transport properties, including complex magnetic behaviors, a large negative MR, and a nonlinear AHE. Essentially, Eu$_{11}$Zn$_6$As$_{12}$ is an antiferromagnet with uncompensated FM component. Three magnetic transitions are consistently identified through the magnetization, specific heat, and resistivity measurements: FM fluctuations around 29 K ($T_\mathrm{F}$), AFM ordering at 22 K ($T_\mathrm{N}$), and an AFM-like transition at 9 K ($T^*$). The resistivity of Eu$_{11}$Zn$_6$As$_{12}$ is one order of magnitude higher than that of the isostructural compound Eu$_{11}$Cd$_6$Sb$_{12}$, and the strong FM fluctuations above $T_\mathrm{N}$ result in a prominent resistivity peak around 18 K. Moreover, a resistivity hysteresis is observed at low temperatures ($<5$ K) due to the remnant FM component. The resistivity peak around $T_\mathrm{N}$ diminishes rapidly in increasing magnetic fields, leading to a maximum MR effect of $-190\%$ at 5 T. Our analysis of the field-dependent Hall resistivity suggests a carrier concentration of $1.2\times 10^{20}$ cm$^{-3}$, which may result from the commonly observed cation defects in Zintl phases or the mixed valence of Eu$^{2+}$ and Eu$^{3+}$. A nonlinear AHE contribution occurs around $T_\mathrm{F}$ and $T_\mathrm{N}$, similar to the cases in CaAl$_2$Si$_2$-type Eu$M_2X_2$ compounds~\cite{xuUnconventionalTransverseTransport2021,Yi2023,Singh2024,wangColossalMagnetoresistanceMixed2021}, which could be attributed to the nonzero spin chirality due to the noncollinear spins of Eu sites.

Our study demonstrates that the magnetism and charge transport properties of Eu$_{11}$Zn$_6$As$_{12}$ are distinct from those of other Eu-containing 11-6-12 phases, indicating that even sister materials with the same structure, specifically narrow-gap rare-earth Zintl phases, can exhibit substantially different physical properties. Therefore, the structural diversity of Zintl phases presents an exceptional platform for exploring the interplay between charge transport and magnetic ordering through tuning the composition. In addition, further experiments are required to elucidate the unusual properties of Eu$_{11}$Zn$_6$As$_{12}$, such as unraveling the complex spin structure and determining the true valence state of the Eu ions.

\begin{acknowledgments}
This work was supported by the National Natural Science Foundation of China (Grants No. 12204094 and No. 12325401), the Natural Science Foundation of Jiangsu Province (Grant No. BK20220796), the Start-up Research Fund of Southeast University (Grant No. RF1028623289), the Interdisciplinary program of Wuhan National High Magnetic Field Center (WHMFC) at Huazhong University of Science and Technology (Grant No. WHMFC202205), and the Big Data Computing Center of Southeast University.
\end{acknowledgments}

\appendix*
\section{Magnetization}
\begin{figure*}
	\includegraphics[width=0.9\textwidth]{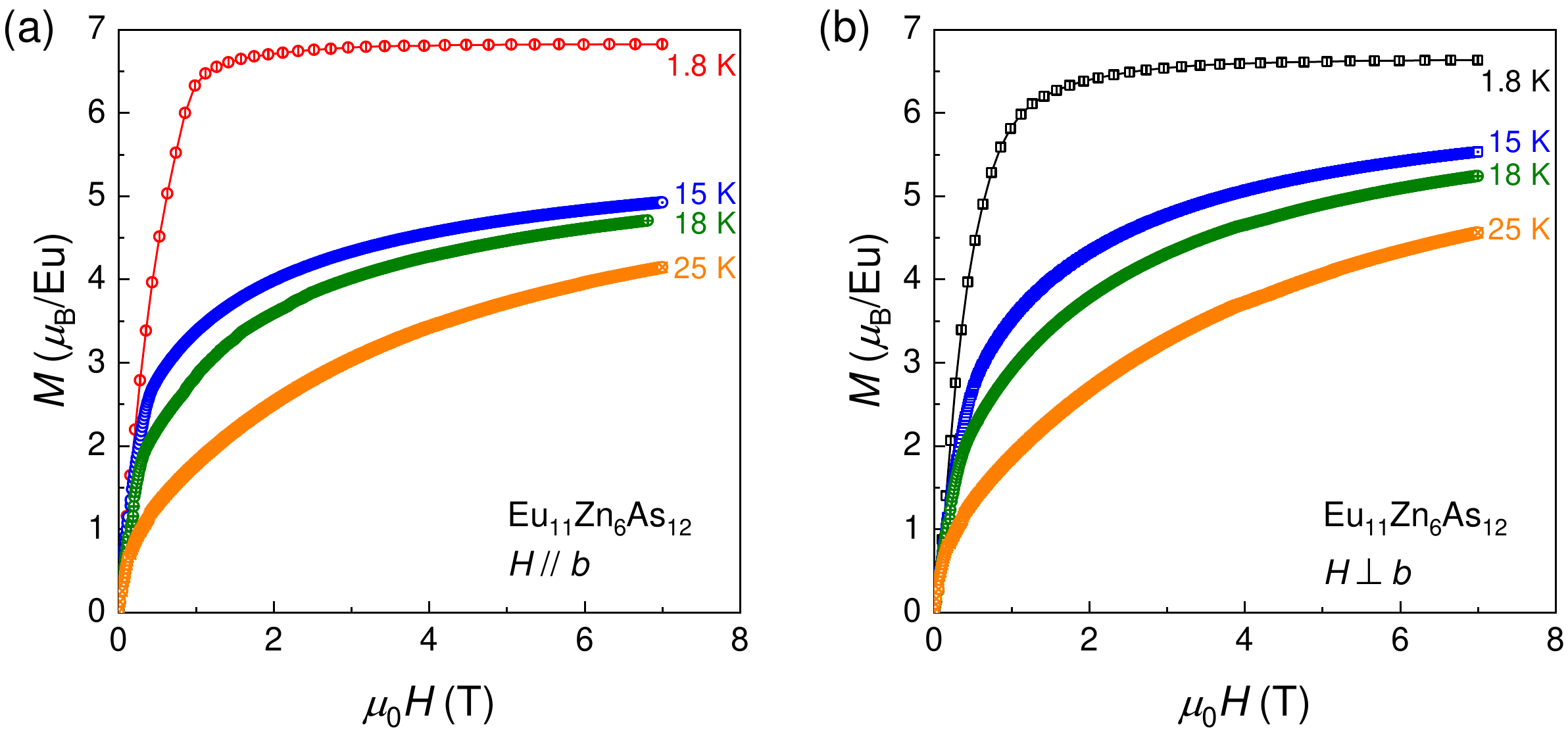}
	\caption{Anisotropic magnetization of Eu$_{11}$Zn$_6$As$_{12}$ at 1.8, 15, 18, and 25 K, as a function of fields (a) along  and (b) perpendicular to the $b$ axis.}	
	
\label{FA}
\end{figure*}

Magnetization as a function of field $M(H)$, with fields applied along and perpendicular to the $b$ axis, is shown in Figs.~\ref{FA}(a) and (b), indicating a weak magnetic anisotropy.

\bibliography{reference}
\end{document}